\RequirePackage[hyphens]{url}

\documentclass[11pt,letterpaper]{article}

\usepackage{graphicx}
\usepackage{setspace}
\usepackage{longtable}
\usepackage{booktabs}

\usepackage{babel}
\usepackage{lipsum} 

\title{Gender Bias in Password Managers}
\author{Jeff Yan and Dearbhla McCabe \\ University of Strathclyde, UK \\ Email:\{first.last@strath.ac.uk\}}

\date{}
\begin{document}
\maketitle
\begin{abstract}

For the first time, we report gender bias in people's choice and use of password managers, through a semi-structured interview ($n=18$) and a questionnaire-based survey ($n=200$, conducted `in the wild'). Not only do women and men prefer different types of password managers, but software features that women and men frequently use also differ. These differences are statistically significant. The factors that women and men consider the most important or influential in choosing their password managers differ, too. Choice of convenience and brand are on the top of the women's consideration, whereas security and the number of features top the list for men. This difference is statistically significant. 

\end{abstract}













\setstretch{1}
\section{Introduction}

Usable security has gained traction in the past 25 years,  in both academia and industry, since the seminal `Why johnny can’t encrypt' study \cite{alma99} which Whitten and Tygar conducted in 1998 helped launch the field. 
Gender is usually collected among demographics in usable security studies.
However, in this fast-moving community, 
there has been little research concentrating on gender aspects of 
security, 
except for a few exceptions. 
For example, Moore and Anderson \cite{ross08} 
reported an empirical study where anti-phishing advice offered by some banks were found easier 
for men than women to follow. 
In the research of deception\footnote{Deception is not only the basic problem underlying security and cybercrime, but is central to human behaviour.} deterrence in socio-technical systems led by Cambridge University, gender biased deception in poker \cite{yan16} was 
studied. 

Gender and diversity matter for security. 
For a long time, it used to be young geeky men writing software or engineering secure systems for male geeks. Not any more. The user base of security technologies has been greatly diversified, and this has wide implications. First, as widely known and appreciated nowadays, poor usability could lead to security failures. Second, engineers without gender awareness/knowledge will likely create products that lead to gender discrimination, either explicitly or implicitly, and put some of the population at a disadvantage. 
Gender inequality, if not properly addressed, could cause security failures, too. For example, people who are or feel discriminated against will feel unfair and resentment, it reduces their compliance of security policies and mechanisms but increases their propensity to circumvent them. 
On the other hand, gender studies have been a prominent subject in social sciences for long. In computer science, gender has informed the field of HCI \cite{Cassell02}. 
We believe that studying gender will 
offer valuable and even unique insights to inform security, too. 

In this article, we report the first 
study that examines gender bias in password managers, which are a popular security tool for people to use passwords more effectively. 

We all have too many user accounts and passwords, but limited human memory. It has been known for long that mnemonic passwords achieve good security and memorability simultaneously \cite{yan99}. However, remembering a number of unique and strong passwords, as well as their associated accounts and services, is still demanding for everyone. 

Password managers were invented to serve a number of goals, notably, reducing people's cognitive burden of remembering usernames, passwords and their associated services;  improving security by encouraging or even generating strong passwords and by reducing password reuse, a poor but common practice. 

Numerous products are available for password managers on the market. And there is a substantial and growing body of academic literature on password manager research. However, little is known about gender aspects of password managers. 

We ask two main research questions: 
Do men and women choose password managers differently?
Do men and women use them differently, too?

To answer these questions, we conducted both qualitative and quantitative experiments. 
We find that women and men differ significantly not only in their choice and use of password managers, but also in their decision making, such as how they have made their choice.


\section{Methodology}

\subsection{Theory and hypothesis}

Many different password managers are available for people to choose, and they can be classified into three categories:

\begin{itemize}
\item Browser built-ins, which are built into browsers, provided by Google Chrome, Microsoft Edge, Internet Explorer, Firefox, etc.

\item OS built-ins, which are built into operating systems, for example, Apple's keychain and Samsung Pass.
\item Standalone ones, which are not built into browsers or operating systems and must be separately installed, for example, Dashlane, 1Password, LastPass, KeePass, RoboForm, to name a few. Typically, these are from third-party vendors.

\end{itemize} 
With regard to where passwords are stored, each category 
might support local storage, cloud storage, or both.

We postulate that when people choose their password managers, they will consider a range of factors for their decision making, namely: 

\begin{itemize}
\item Brand: a product's brand reputation;

\item Security: whether a product provides good security; 

\item Usability: whether a product is easy to use; 

\item Features: whether a product is feature-rich; 

\item Choice of convenience: the usual `shortcut route' approach for decisions in life. Clearly this is not about usability. 

\item Location of password storage: are passwords stored locally, on the cloud, or both?

\end{itemize}

We have three main hypotheses:

1) Women and men differ in their choices of password managers. 

2) Women and men differ in their most influencing factors for password manager choices.

3) Women and men differ in their use of password managers.

\subsection{Experiments}

We design a two-stage study: a semi-structured interview as a pilot, followed by a main study, 
each with different participants. Interviews cannot easily scale to hundreds of participants. Therefore, we design our main study as a questionnaire-based survey, aiming for statistically significant results. 

We intend to use the pilot to test our theory quickly, and if needed make necessary adjustments accordingly. We expect the pilot will inform our questionnaire design, too. This way, we will speed up our research in a cost-effective way and without losing rigor.

Both the pilot and the survey aim to explore people's real-life experience in choosing and using their password managers. 
For both, 
we explicitly invite only password manager users (i.e. people who use or have used one), but exclude non-users. 

Both the pilot and the survey collect demographics including age, gender, education level, discipline, security experience. Discipline allows one choice of two: STEM (Science, Technology, Engineering, Mathematics) vs others.

\subsubsection{Pilot study}

We aim for semi-structured interviews, 
conducted in a relaxed manner. 
The more relaxed an interview is, the more likely a participant is to reveal honest opinions rather than what they believe we want to find out. 

We mainly explore two themes: 1) what password manager do you use? and 2) why do you choose to use it? Further questions will be prompted based on a participant's answers to previous questions, keeping in line with the framework of our theory, but the goal is to ensure the participant feels all power dynamics are vanished, keeping the interview conversational, bouncing off with our friendly interviewer. For this reason, the interviewer may share self experiences, while ensuring that the interview remains unbiased.

Our interview explores what factors in our framework a participant regards as the most important for selecting their password managers. However, the questions we ask will mostly based on ongoing discussions. We encourage each participant to disagree and contribute their own suggestion.

Finally, we conduct a careful thematic analysis of interview transcripts to derive our results.

The pilot has served us well. It has led to interesting and promising observations, which most importantly provide the first set of evidence to support our theory and hypotheses. Detailed results are useful but the sample set is not representative, and thus we omit them in this article. We notice that some fine prior studies on password managers were based solely on interviewing a small set of participants.

For reference, we include a summary of our demographic data here. 
In total, 20 participants were recruited; two had to withdraw due to schedule conflicts. 9 females and 9 males completed the interview. 12 participants came from a STEM discipline; 16 participants had a degree or higher education. 15 participants were aged between 18-24, the remaining three one each in the groups of 25-34, 35-44 and 45-54. 10 participants had no or little security experience, 1 participants had 1-3 years, and 7 had 4+ years.

\subsubsection{Survey questionnaire}

Our questionnaire has 21 questions in total, aiming for a quick survey to be completed in 10-15 minutes on average.

We set to solicit our participants' truthful answers. Deliberately, we do not start our 
questionnaire with demographics questions, in order to prevent them from consciously or sub-consciously second-guessing
answers that they think we want, e.g. by associating their demographic data with some answers. Instead, we open with two questions, asking (Q1) their experience in cyber security (in years), and (Q2) how good they feel about their understanding of the best cybersecurity practices.

Next four questions (Q3-6) ask education level, discipline, age, and gender sequentially. 

Q7 asks what types of password managers they use. This is a multiple-choice question, and three categories (browser built-ins, OS built-ins and standalone) are given. To avoid confusion, we use `Standalone' instead of `third party'. To help each participant to decide what types of password managers they use, we give many examples in each category in the survey. 

Next ten questions (Q8-17) correspond to the six factors in our framework. For each factor, we design one to three  questions to probe a participant's understanding and opinion. For these questions, participants are expected to answer on a five-point scale ranging from 1 `Strongly Agree' to 5 `Strongly Disagree'.

Q18-19 explore people's risk perception, although it is not part of our framework. 


Q20 explores common features of 
password managers that people use frequently.  
Password storage and autofill\footnote{Fill passwords automatically so that a user does not have to enter it manually.} are two basic features which everybody uses,
even if they does not realise it. Therefore, we are interested only in the use of additional features such as password generation, easy password reset, synchronisation between devices, store sensitive data
and VPN access. This is a multiple-choice question, and a participant can chose one or all answers. 

The decision on these five features is based on our experience, a fine systematization paper \cite{simmons2021systematization}, as well as the output generated from our pilot. VPN access was not on our list initially. 
However, some password managers such as Dashlane, LastPass, NordPass and Kaspersky either offer VPN as a feature or are available with a VPN in a bundle. We have a strong impression from the pilot that VPN access was widely regarded as a common feature among our participants. Therefore, we include it on the list.


Our last question (Q21) asks a participant to 
pick only one out of six candidate factors \textit{brand, security, usability, the number of features, location of password storage, choice of convenience} as the most important factor in choosing their password managers

Two questions (Q2 on self-reported confidence in security knowledge, Q18 on risk perception) also serve the purpose of sanity checking for our survey. 

\subsection{Research ethics}

We obtained full approval from our Ethics Committee at Strathclyde before our study started. For the pilot and the survey, we sought each participant's informed consent before enrolling them. Their participation was voluntary and anonymous, and we made it explicit and clear that they could drop out at any stage.




\section{Results}

\subsection{Demographics}

We collected two hundred completed responses (100 male, 100 female) for data analysis. 
We followed an \textit{`in the wild'} approach in the hopes to gain as much perspective as possible from a wide range of people, reflective of the modern population, except for that a participant must be a password manager user, and at the age of 18 or over. Our online questionnaire was written in English, and recruit advertised by social media/networks. Presumably, the majority of participants are based in the UK. We did not collect race and country data, though.

\par

Table \ref{tab:dmg} summarises participant demographics.
In this sample set, female participants and males do not differ significantly in the distribution of security experience, discipline, age or education level. Our analysis does not return any statistically significant effect.

A substantial portion of our participants did not declare extensive security experience or a higher education. This is different from most, if not all, previous studies on password manager users. 
It is difficult to be certain whether this is a truly representative sample of password manager users, but for sure it is more representative than those in prior studies. 

Gender wise, our questionnaire (as well as our pilot) was open to non-binary participants, but unfortunately we received no responses from that demographic.
Alongside this, our analysis will look at gender from a binary perspective, but use female/male as a social concept, rather than strictly biological.

\begin{table}[h!]
		\centering
		\begin{tabular}{p{10em}rll}
			\multicolumn{2}{r}{}      & \multicolumn{1}{p{3em}}{Male} & \multicolumn{1}{l}{Female} \\
			\hline
			Experience in  & \multicolumn{1}{p{10em}}{Fewer than 1 year} & 22          & 28 \\
			cyber security & \multicolumn{1}{p{10em}}{1-3 years} & 40          & 30 \\
			\multicolumn{1}{r}{} & \multicolumn{1}{p{10em}}{4+ years} & 17          & 19 \\
			\multicolumn{1}{r}{} & \multicolumn{1}{p{10em}}{None/little} & 21          & 23 \\ \\
			Discipline & \multicolumn{1}{p{10em}}{STEM} & 52          & 54 \\
			\multicolumn{1}{r}{} & \multicolumn{1}{p{10em}}{Others} & 48          & 46 \\ \\
			Age & \multicolumn{1}{p{10em}}{18-24 years} & 32          & 31 \\
			\multicolumn{1}{r}{} & \multicolumn{1}{p{10em}}{25-34 years} & 18          & 17 \\
			\multicolumn{1}{r}{} & \multicolumn{1}{p{10em}}{35-44 years} & 25          & 24 \\
			\multicolumn{1}{r}{} & \multicolumn{1}{p{10em}}{45-54 years} & 15          & 17 \\
			\multicolumn{1}{r}{} & \multicolumn{1}{p{10em}}{55-64 years} & 8           & 8 \\
			\multicolumn{1}{r}{} & \multicolumn{1}{p{10em}}{65+ years} & 2           & 3 \\ \\
			Education & \multicolumn{1}{p{10em}}{High school} & 22          & 11 \\
			\multicolumn{1}{r}{} & \multicolumn{1}{p{10em}}{Degree} & 57          & 69 \\
			\multicolumn{1}{r}{} & \multicolumn{1}{p{10em}}{Masters or above} & 21          & 20  
		\end{tabular}%
		\caption{\label{tab:dmg}Demographics summary (participants: 100 males and 100 females)}
\end{table}%


\subsection{Password manager choices}

Table \ref{tab:pm_gender} shows the choices of password managers by types (the results from Q7). Specifically, 73\% males  use standalone password managers, but 97\% females stay away from them altogether. Instead, 66\% females use password managers built into browsers, and 49\% females use password managers built into operating systems. Females and males differ significantly in their password manager choices (Fisher-Freeman-Halton test, $p < .001$). 

\begin{table}[tbp]
	\centering

	\begin{tabular}{lrll}
		\multicolumn{1}{r}{} &             &             &  \\
		\multicolumn{2}{p{15em}} {\textbf{Password Manager Type}}     & \multicolumn{1}{p{7.43em}}{Male} & \multicolumn{1}{p{7.715em}}{Female} \\
		\hline
		OS built-ins    &             &          16   &  49\\
		Browser built-ins &             &            29  &  66 \\
		Standalone &             &          73   &  3\\
	\end{tabular}%
		\caption{\label{tab:pm_gender}Password manager choices by types}
\end{table}%

Our question allows multiple choices, and each participant can truthfully report 
all password managers which they use. As each column in Table \ref{tab:pm_gender} adds up to over 100, 
clearly some males and females use more than one type of password managers. Do females and males differ
in their password manager combinations? 
Table \ref{tab:pm_pt}
provides a participant classification by 
password manager combinations.
Each row defines a possible combination, 
and all combinations 
(rows) are mutually exclusive.

\begin{table}[h!]
	\centering

	\begin{tabular}{lrll}
		\multicolumn{1}{r}{} &             &             &  \\
		
		\multicolumn{2}{p{15em}}{
		Password manager combinations}      & \multicolumn{1}{p{7.43em}}{Male} & \multicolumn{1}{p{7.715em}}{Female} \\
		\hline
		O + B + S    &             &          1   &  1\\
		B + S &             &            4  &  0 \\
		S &             &          68   &  2\\
		O + B &             &          12   &  16\\
		O + S &             &          0   &  0\\
		B &             &          12  &  49\\
		O &             &          3   &  32 \\ 
		\hline
		\multicolumn{2}{p{10em}}{Total} & 100           & 100\\
	\end{tabular}%
				\caption{\label{tab:pm_pt} 
		Password manager choices by combination. O: those built into operating systems; B: those built into browsers; S: standalone ones. 
		}
\end{table}%

Among our participants, females and males still differ greatly in their password manager combinations. 
97\% females use password mangers built into browsers, or those built into operating systems, or both. Only 3\% females use standalone ones. However, 68\% males use only standalone password mangers. 

While some females and males use two or three types of password managers together, most of them use only one type. 80\% males use either standalone ones or those built into browsers, whereas 81\% females use either browser built-ins or those built into operating systems.


\subsection{The most important factor for choices}

Table \ref{tab:mif} summarises the factors which our participants consider the most important in choosing their password managers (the results from Q21). In this regard, female and male participants differ significantly 
(Fisher-Freeman-Halton test, $p < .001$).

Specifically, choice of convenience is the top influencing factor for females (with 48\% votes), and brand the second (35\% votes); whereas only 9\% males consider choice of convenience the most important factor, and only 6\% males consider brand the most important. 

For males, security is the first priority (with 38\% votes), and the number of features available a close second (35\% votes); whereas only 8\% females consider security the most important factor, and 3\% females consider the number of features the most important factor. 


Location of password storage has got the least votes from both females (4\%) and males (2\%). Usability has got the second least votes from both females (8\%) and males (4\%).

\begin{table}[h!]
	\centering

	\begin{tabular}{lrll}
		\multicolumn{1}{r}{} &             &             &  \\
	
		\multicolumn{2}{p{15em}}{\textbf{Most Important Factor}}  & \multicolumn{1}{p{7.43em}}{Male} & \multicolumn{1}{p{7.715em}}{Female} \\
		\hline
		Brand    &             &          6   &  35\\
		Choice of convenience &             &            9  &  48 \\
		Location of password storage &             &          4   &  2\\
		Number of Features &             &          35   &  3\\
		Security &             &          38   &  8\\
		Usability &             &          8   &  4\\ 
		\hline
		\multicolumn{2}{p{15em}}{
		Maximum votes per factor} & 100           & 100\\
	\end{tabular}%
		\caption{\label{tab:mif}The most important factor in choosing password managers}
\end{table}%

We use logistic regression analyses to examine the association between the most important factor for choices and participant characteristics. 
Gender has a statistically significant and positive effect on the most important choice factor, controlling for possible confounding factors including discipline, security experience, education level, age and self-reported confidence in security knowledge ($p < .001$). Discipline does not have a significant effect. Therefore, gender is the only variable that predicates the most 
important factor for password manager choices.

\subsection{Frequently used features}

Table \ref{tab:ffu} summarises five frequently used features, 
with votes from our participants (corresponding to Q20). 
Clearly, much more men use each of these features than women do (all statistically significant, $p < .001$).

\begin{table}[!ht]
	\centering

	\begin{tabular}{lrll}
		\multicolumn{1}{r}{} &             &             &  \\
			\multicolumn{2}{p{15em}}{\textbf{Features}}      & \multicolumn{1}{p{7.43em}}{Male} & \multicolumn{1}{p{7.715em}}{Female} \\
			\hline 
		Password generation &             &          52   &  16\\
		Easy password reset &             &        51     & 16  \\
		\multicolumn{2}{p{15em}}{Synchronisation between devices} &    57         &  20\\
		\multicolumn{2}{p{10em}}{Store sensitive data} &       32      & 8  \\
		\multicolumn{2}{p{10em}}{VPN access} &         49    &  6\\	
		\hline
		
		\multicolumn{2}{p{15em}}{
		Maximum votes per feature} & 100           & 100\\
	\end{tabular}%
		\caption{\label{tab:ffu}Use frequency of extra features 
		(in addition to password storage and autofill)}
\end{table}%

\newpage


How many features, in additional to password storage and autofill, do men and women frequently use, 
respectively? To answer this interesting question, Table \ref{tab:c_by_fno} provides a participant classification by the number of extra features they frequently use (from 0 to 5). All rows are mutually exclusive. 

We observe some interesting patterns. 
First, the majority of female participants (68\%) use no extra features frequently at all, whereas most males (83\%) use at least two.

Second, the number of females monotonically decreases as the number of features used increases. That is, the more features, the fewer females use them frequently. 

Third, it is not the case that the more features, the more males use them all. A large portion of male participants (68\%) use two or three extra features frequently, and only 15\%  males use four or more. 

The contrast between female and male participants becomes apparent when we plot Table \ref{tab:c_by_fno} data in Figure 1.

\begin{table}[h!]
	\centering

	\begin{tabular}{lrll}
		\multicolumn{1}{r}{} &             &             &  \\
			\multicolumn{2}{p{15em}}{\textbf{\# Extra features used}}      & \multicolumn{1}{p{7.43em}}{Male} & \multicolumn{1}{p{7.715em}}{Female}\\
			\hline 
		5 (all)  &             &          6   &  1 \\
		4 only &             &        9    & 3   \\
		3  only &    & 29       &  6 \\
		2 only &      & 39  & 9  \\
		1 only &     &  10 & 13  \\
		0  &       &  7    &  68 \\	
		\hline
		\multicolumn{2}{p{15em}}{Total 
	    } & 100           & 100\\
	
	\end{tabular}%
		\caption{\label{tab:c_by_fno}Classifying participants by the number of extra features they frequently use}
\end{table}%

\begin{figure}[!ht]
\centering
\includegraphics[scale=0.8]{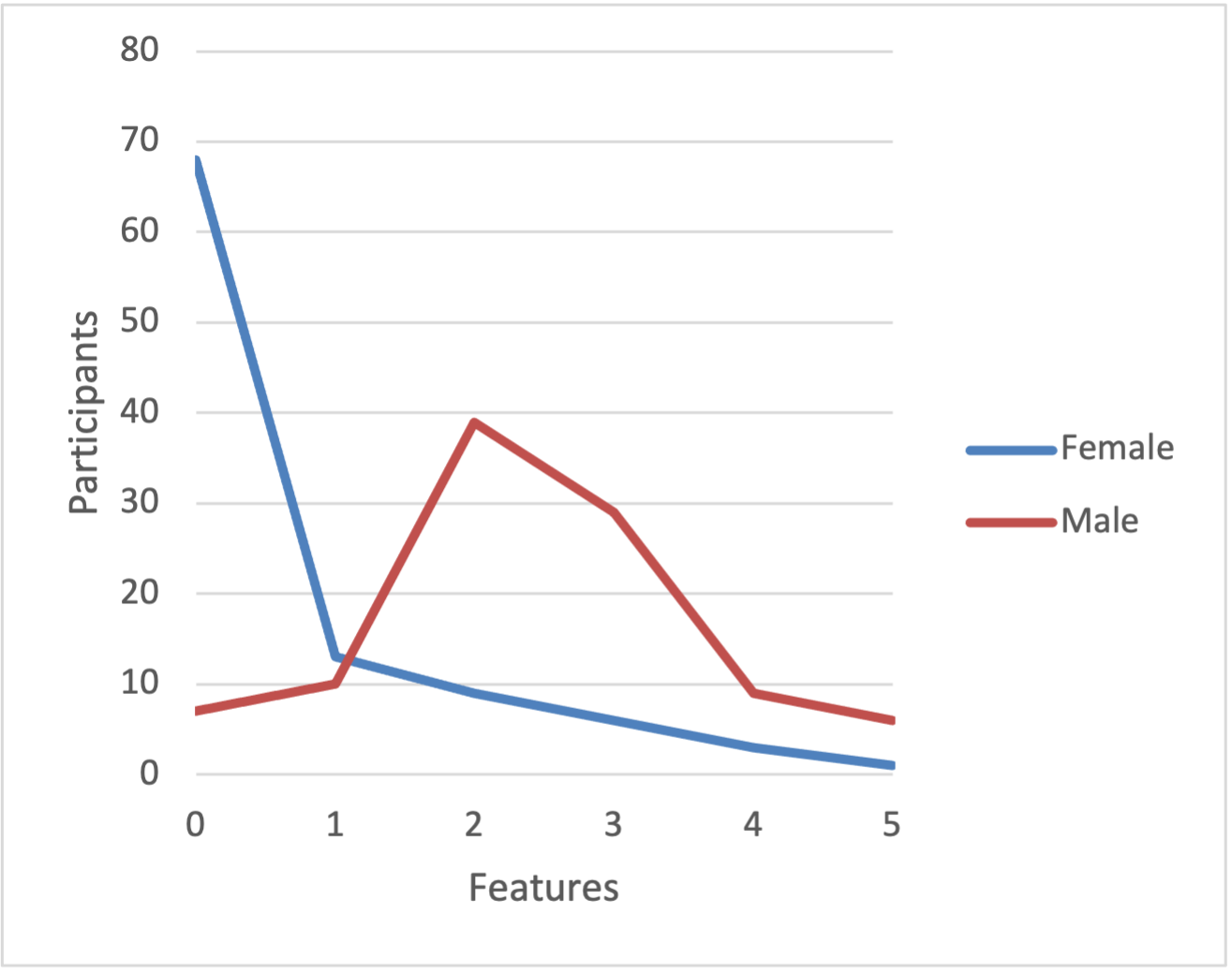}
\caption{\label{fig:featuresused}Use of extra features: interesting contrasts by gender.}
\end{figure}

\newpage

\subsection{Individual factors}

\subsubsection*{Brand}

\noindent
Q8 \textit{`When selecting a password manager, I look for a brand with a strong
reputation in security.'}: 89\% males and 22\% females agree or strongly agree, whereas 62\% females and only 6\% males disagree or strongly disagree.

\noindent
Q9 \textit{`When selecting a password manager, I look for a brand with a strong
reputation overall, not necessarily security related.'}:
73\% females and 22\% males agree or strongly agree, whereas 67\% males and only 7\% females disagree or strongly disagree.

Therefore, both males and females care about brand. However, 
when selecting a password manager, for most males, a strong brand in security trumps a brand with a strong reputation overall. For most females, it is the other way around. These differences are statistically significant (Fisher-Freeman-Halton test, $p < .001$).
This contributes to explaining why males prefer standalone password managers, and females like those built into big-name operating systems or browsers.

\subsubsection*{Choice of convenience}

\noindent
Q10 \textit{`I chose my password manager, simply because it was readily available on my device.'}: 93\% females and 24\% males agree or strongly agree, whereas 66\% males and only 2\% females disagree or strongly disagree.

\noindent
Q11 \textit{`I spent time researching password managers before making my choice.'}: 61\% males and only 10\% females agree or strongly agree, whereas 86\% females  and  15\% males disagree or strongly disagree.

\noindent
Q12 \textit{`The initial inconvenience of downloading a password manager does impede my final choice.'}:  80\% females and 22\% males agree or strongly agree, whereas     64\% males and 10\% females disagree or strongly disagree.

For each of these questions, the response distribution differs significantly for female and male participants (Fisher-Freeman-Halton test, $p < .001$). Clearly, choice of convenience has a significant impact on women, but not on men.

\subsubsection*{Location of password storage}
\noindent
 Q13 \textit{`I would prefer a password manager to store my password credentials'}: 49\% females prefer local storage, 13\% prefer cloud storage, and 38\% are ok with either local or cloud storage. 28\% males prefer cloud storage, 24\% prefer local storage, and 48\% choose either local or cloud storage. The response distribution differs significantly for female and male participants (Fisher-Freeman-Halton test, $p < .001$).
    
\subsubsection*{Number of features}
\noindent
    Q14 \textit{`The more features I get from a password manager, the better.'}: 16\% females and 65\% males agree or strongly agree, whereas     15\% males and 66\% females disagree or strongly disagree. 
\noindent
    Q15 \textit{`I use a password manager only for the basic functions, and I do not require anything else.'}:
    88\% females and 26\% males agree or strongly agree, whereas     64\% males and 8\% females disagree or strongly disagree.
    
    For both questions, the response distribution differs significantly for female and male participants (Fisher-Freeman-Halton test, $p < .001$). Clearly, features have a significant impact on males, but not on females. 
    
\subsubsection*{Security}
\noindent
    Q16 \textit{`When selecting password managers, I care and look into the level of security offered.'}:
    18\% females and 86\% males agree or strongly agree, whereas     8\% males and 69\% females disagree or strongly disagree.
    Most male participants care about security, but most females do not seem so. The response distribution differs significantly for females and males. (Fisher-Freeman-Halton test, $p < .001$).

\subsubsection*{Usability}        
\noindent   
    Q17 \textit{`The easier a password manager is to use, the more likely I will choose it.'}: 
    53\% females and 59\% males agree or strongly agree, whereas     17\% males and 17\% females disagree or strongly disagree. The remaining 24\% males and 30\% females stay neutral. 
    
    The response distribution 
    is not significantly different for female and male participants (Fisher-Freeman-Halton test, p = .138). This is one of the rare factors which men and women have a similar opinion.  

\subsection{Risk perception}    
\noindent
    Q18 \textit{`The thought of being hacked online leaves me feeling anxious.'}:
    90\% females and 56\% males agree or strongly agree, whereas     22\% males and only 6\% females disagree or strongly disagree. The response distribution is significantly different for males and females (Fisher-Freeman-Halton test, $p < .001$). That is, women are more worried about being hacked online than men are. This is in agreement with the conventional wisdom that women have a higher level of anxiety with cyber security.
    
    Q19 \textit{`Storing my passwords on a password manager reduces my anxiety of being hacked.'}: 
    20\% females and 59\% males agree or strongly agree, whereas 7\% males and 69\% females disagree or strongly disagree. The response distribution is significantly different for females and males (Fisher-Freeman-Halton test, $p < .001$). More than half males believe that password managers help with their anxiety reduction, but nearly 70\% females disagree. 

\subsection{Self-reported confidence in security knowledge}
\noindent
Q2 \textit{`I feel I have a good understanding of the best cybersecurity practices.'}:
42\% female participants and 60\% males agree or strongly agree, whereas 10\% males and 22\% females disagree or strongly disagree. 30\% males and 36\% females stay neutral. The response distribution is statistically different for females and males (Fisher-Freeman-Halton test, $p = .011$). That is, men in general are more confident than women in their security knowledge. This is in agreement with conventional wisdom.

\section{Discussions}
Comparing each gender's choices of the most important factor with their responses to other questions in the survey, 
we can see that they corroborate and complement each other.

Women voted choice of convenience as their most important factor. Q10, Q11 and Q12 clearly indicate that women want to avoid researching and downloading password managers, and they are willing to use one that is already readily available.

Women voted on brand reputation as the second, but they care a brand's reputation overall more than in security (Q8-9).
This is also consistent with women placing security the second last in importance.

Women's password manager choices also support our analysis. A majority used the password managers provided by their operating systems or browsers. This indicates a willingness to use what was readily available. Also, these
password managers tend to be created by well-known major technology brands, which have a high reputation in many areas of technology and should have influenced female decision making, too. 

Men voted security as their most important factor. Q16 clearly indicates that they 
care a lot about the level of security offered in their password manager.Q9 and Q8 shows that a majority of men care much more about a brand's reputation in security than its overall reputation. Altogether, brand does not have a significant impact on men. 

The number of features was second in men's choice as most important factor. Q14, Q15 and Q20 show they want and indeed use more than just the basic functions of a password manager. Men clearly place high value in features offered in a password manager, since their answers are consistent throughout the questionnaire in all the questions related to features.

Women, in contrast, did not place the number of features high.
This is consistent amongst the rest of the survey, with women in Q15 only wanting the basic functions in a password manager, and only utilising either one or no extra features offered in a password manager in Q20. \par

On the other hand, two sanity-checking questions (Q2 on self-reported confidence in security knowledge, and Q18 on risk perception) both return affirmative results in agreement with the conventional wisdom. We detect no abnormality in our 
data collection. Therefore, 
we believe that our survey results are robust.

Moreover, we observe some interesting phenomena. First, 
Q17 showed that both genders looked for good usability. However, both did not place usability high in their most important factor. 
A possible and speculative explanation is that they consider current password managers already provide reasonable or good enough usability.

Second, nearly 60\% male participants felt their anxiety was reduced by using a password manager, whereas nearly 70\% of women did not (Q19). This is likely because men value security more in their decision-making process than women do, 
and therefore understand more a password manager's role in security. 

It is interesting to figure out solid explanations of these observations, but beyond the scope of our study. Instead, we leave them for future research.

\section{What do we inform developers of?}

To attract male users to browser or OS built-ins, password manager developers could consider increasing the number of features their products offer. However, features help, but only to an extent. 
As we have observed, the number of frequent users among men starts to significantly drop when the number of extra features reaches four (see Figure 1).

It could also help to attract male users, if browser and OS built-ins increase their security, both for real and for user perception. 

To attract women users, it may be a useful strategy for standalone password managers to get pre-installed onto devices or even browsers. Pre-installation will increase ‘choice of convenience’ for women by reducing their reluctance and burden of search, research, downloading and installation. However, the brand reputations of standalone 
products do not appear to work for women (effectively). In this regard, brand building may help. 

Some further thoughts on features turn out to be beneficial for both men and women. It is not that the more features, the better. This holds true for both men and women. However, there is a significant difference, namely, men tend to want all features they consider useful, but women tend to use only the basic ones. The lesson for the developers is therefore the following: getting the basic features done brilliantly, but giving the men all the features they consider essential and useful, and getting these features done really well.

Nonetheless, the best way forward appears to be for a big-name player to acquire a decent standalone solution, get it pre-installed or integrated into a browser or OS, and get it re-branded with the big name. This way, it will increase attractions to both men and women users at the same time. {\bf If this strategic suggestion is adopted, the acquisition beneficiaries please make a donation to support our research.}
 
In the competition for men and women users, what new equilibrium will be established for password managers? This is an interesting open question to observe and ponder.

 \section{Related work}

Password managers have been extensively studied, including their security, usability and adoption challenges.  Our literature review will be brief, and pay attention to 1) major studies in password manager research, and 2) the latest development since 2019, for two reasons. First, Chaudhary et al. \cite{bigsurvey2019} conducted a systematic literature review on this topic in 2019, covering over 30 papers. Second, 
prior art in the literature, albeit fine work, collectively tells us little about gender aspects of password managers.

{\bf Security.} Major studies include vulnerabilities in password managers built into browsers \cite{Li2014}\cite{oesch2020then}, vulnerabilities in Autofill \cite{silver2014password}\cite{oesch2021emperor}, and the impact of password managers on users' real-life password strength and reuse \cite{Lyastani2018}.

{\bf Usability.} 
Chiasson et al. \cite{chiasson2006usability}  reported usability issues in two early password managers, some of which 
could lead to security problems.
Seiler-Hwang et al. \cite{SeilerHwang2019} studied the usability of mobile password managers.
A fine systematization paper by Simmons et al. \cite{simmons2021systematization} 
 suggested that the literature has not investigated all use cases yet.
Huaman et al. \cite{huaman2022they} investigated how to make password managers better interact with the Web.
Oesch et al. \cite{oesch2022basically} 
reported an observational interviews with 32 users, aiming to explain their use of password managers via the grounded theory approach.

{\bf Adoption.} 
Alkaldi and Renaud \cite{karen19} investigated how effective the self-determination theory, originated in psychology, was to encourage adoption of password managers.
Pearman et al. \cite{pearman2019people} reported 
a semi-structured interview study with 30 participants, including both password manager users and non-users. 
Ray et al. \cite{ray2021older} replicated the study of \cite{pearman2019people} on senior people.

\section{Conclusion}\label{conc}

We have observed some significant differences in the choice and use of password managers for men and women.

Most men use standalone password managers, but nearly all women except some exceptions stay away from them altogether. Instead, most women use password mangers built into browsers, or those built into operating systems, or both. While some women and men use more than one type of password managers at the same time, most of them use only one.

In choosing password managers, women tend to rely more on choice of convenience and brand reputation, whereas men tend to rely more on security and the number of software features offered. Human decision is a complex procedure. It might well be the case that many factors are considered, with a different weight for each, before a choice is made. We have chosen to understand the most influencing factor, but leave the exact decision-making process for future research. 

Women and men differ significantly in their use of password managers, too. In additional to basic features such as password storage and autofill, most women (68\%) use no extra features frequently, whereas most men (83\%)
use at least two. `The more features, the better' is untrue for both men and women. 
The number of women users monotonically decreases as the number
of features increases. And a large portion of men (68\%) use only two or three extra
features frequently.

We have also discussed the implications of our findings for password manager developers. 
We hope this study offers them useful inspirations and food for thoughts, so that better tools will be created to attract a wider range of users. 

\section*{Acknowledgements}

We sincerely thank all participants of our study. They 
offered valuable insights, and made our study possible.





\setstretch{1.0}  

\bibliographystyle{unsrt} 

\bibliography{bibliography}
\clearpage

\end{document}